\documentclass[iop]{emulateapj}

\newcommand{\Kepler}{{\it Kepler}}

\newcommand{\emcee}{{\it emcee}}
\newcommand{\vespa}{\texttt{vespa}}

\usepackage{xcolor}
\usepackage{verbatim}

\newcommand{\thisstar}{WASP-47}
\newcommand{\thisplanet}{WASP-47\,b}
\newcommand{\thissecondplanet}{WASP-47\,\textcolor{black}{e}}
\newcommand{\thisthirdplanet}{WASP-47\,d}

\newcommand{\mearth}{M$_\oplus$}
\newcommand{\rearth}{R$_\oplus$}

\newcommand{\fe}{+0.36}
\newcommand{\ufe}{0.05}
\newcommand{\loggst}{4.34} 
\newcommand{\uloggst}{0.01}
\newcommand{\teff}{5576}
\newcommand{\uteff}{67}

\newcommand{\meh}{[M/H]}

\newcommand{\mst}{1.04}
\newcommand{\umst}{0.08}
\newcommand{\rst}{1.15}
\newcommand{\urst}{0.04}

\newcommand{\ldone}{0.399}
\newcommand{\uldone}{0.020}
\newcommand{\ldtwo}{0.420}
\newcommand{\uldtwo}{0.018}
\newcommand{\rprstb}{0.10186}
\newcommand{\urprstb}{0.00023}
\newcommand{\arstb}{9.715}
\newcommand{\uarstb}{0.050}
\newcommand{\inclb}{89.03}
\newcommand{\uinclb}{0.27}
\newcommand{\impb}{0.164}
\newcommand{\uimpb}{0.045}
\newcommand{\rplb}{12.77}
\newcommand{\urplb}{0.44}
\newcommand{\perplb}{4.1591287}
\newcommand{\uperplb}{0.0000049}
\newcommand{\ttransitb}{2457007.932131}
\newcommand{\uttransitb}{0.000023}
\newcommand{\rprstc}{0.01456}
\newcommand{\urprstc}{0.00024}
\newcommand{\arstc}{3.24}
\newcommand{\uarstc}{0.14}
\newcommand{\inclc}{87.0}
\newcommand{\uinclc}{3.1}
\newcommand{\impc}{0.17}
\newcommand{\uimpc}{0.15}
\newcommand{\rplc}{1.829}
\newcommand{\urplc}{0.070}
\newcommand{\perplc}{0.789597}
\newcommand{\uperplc}{0.000013}
\newcommand{\ttransitc}{2457011.34849}
\newcommand{\uttransitc}{0.00038}
\newcommand{\rprstd}{0.02886}
\newcommand{\urprstd}{0.00047}
\newcommand{\arstd}{16.33}
\newcommand{\uarstd}{0.87}
\newcommand{\incld}{89.36}
\newcommand{\uincld}{0.67}
\newcommand{\impd}{0.18}
\newcommand{\uimpd}{0.16}
\newcommand{\rpld}{3.63}
\newcommand{\urpld}{0.14}
\newcommand{\perpld}{9.03081}
\newcommand{\uperpld}{0.00019}
\newcommand{\ttransitd}{2457006.36927}
\newcommand{\uttransitd}{0.00044}

\newcommand{\mplb}{\textcolor{black}{341}}
\newcommand{\umplb}{\textcolor{black}{$^{+73}_{-55}$}}
\newcommand{\mplc}{\textcolor{black}{\textless 22}}
\newcommand{\umplc}{\textcolor{black}{95\% Confidence}}
\newcommand{\mpld}{\textcolor{black}{15.2}}
\newcommand{\umpld}{\textcolor{black}{$\pm 7$}}

\usepackage{hyperref}
\usepackage{todonotes}

\slugcomment{}

\shorttitle{\thisstar: A Hot Jupiter with Friends}
\shortauthors{Becker et al.}

\begin{document}

\title{WASP-47: A Hot Jupiter System with Two Additional Planets Discovered by K2}

\author{
Juliette C. Becker\altaffilmark{1,2},
Andrew Vanderburg\altaffilmark{2,3},
Fred C. Adams\altaffilmark{1,4},
Saul A. Rappaport\altaffilmark{5},
Hans Martin Schwengeler\altaffilmark{6}}

\email{jcbecker@umich.edu}
\altaffiltext{1}{Astronomy Department, University of Michigan, Ann Arbor, MI 48109, USA}
\altaffiltext{2}{NSF Graduate Research Fellow}
\altaffiltext{3}{Harvard-Smithsonian Center for Astrophysics, Cambridge, MA 02138, USA}
\altaffiltext{4}{Physics Department, University of 
Michigan, Ann Arbor, MI 48109, USA}
\altaffiltext{5}{Physics Department and Kavli Institute for
  Astrophysics and Space Research, Massachusetts Institute of Technology,  Cambridge, MA 02139, USA}
\altaffiltext{6}{Planet Hunter}


\begin{abstract}

Using new data from the K2 mission, we show that \thisstar, a previously known hot Jupiter host, also hosts two additional transiting planets: a Neptune-sized outer planet and a super-Earth inner companion. We measure planetary properties from the K2 light curve and detect transit timing variations, confirming the planetary nature of the outer planet. We performed a large number of numerical simulations to study the dynamical stability of the system and to find the theoretically expected transit timing variations (TTVs). The theoretically predicted TTVs are in good agreement with those observed, and we use the TTVs to determine the masses of two planets, and place a limit on the third. The WASP-47 planetary system is important because companion planets can both be inferred by TTVs and are also detected directly through transit observations. The depth of the hot Jupiter's transits make ground-based TTV measurements possible, and the brightness of the host star makes it amenable for precise radial velocity  measurements.  The system serves as a Rosetta Stone for understanding TTVs as a planet detection technique.

\end{abstract}

\maketitle

\section{Introduction}

Due to their large sizes and short orbital periods, hot Jupiters \citep[roughly Jupiter-mass planets with periods between 0.8 and 6.3 days;][]{nohjs} are among the easiest exoplanets to detect. Both the first exoplanet discovered around a main sequence star \citep{mayor} and the first known transiting exoplanet \citep{charbonneau, henry} were hot Jupiters. Until the launch of the \Kepler\ space telescope in 2009, the majority of known transiting exoplanets were hot Jupiters. Hot Jupiters allow for the determination of many planetary properties, including their core masses \citep{konst} and atmospheres \citep{charbonneausodium}. For these reasons, transiting hot Jupiters were and continue to be the subject of many follow-up studies \citep{kreidberg}.

One such follow-up study is the search for additional planets in the system revealed by small departures from perfect periodicity in the hot Jupiter transit times (called transit timing variations or TTVs). TTVs were predicted \citep{holman, agol} and searched for \citep{steffenagol, gibson}, but very little evidence for TTVs was found until the \Kepler\ mission discovered smaller transiting planets on longer period orbits than the hot Jupiters detected from the ground \citep{kepler9, kepler11}.

The lack of transit timing variations for hot Jupiters implies a dearth of nearby planets in these systems. While systems exist with a known hot Jupiter and a distant ($\gtrsim200$-day period) companion \citep{heather, hjcompanion} or a warm Jupiter (orbital period 6.3 - 15.8 days) and a close-in planet \citep[for example, KOI 191:][]{wjcomp, sanchisojeda}, searches for close-in, companions to hot Jupiters \citep[as in][]{nohjs} have not yet been successful. 

This apparent scarcity supports the idea that hot Jupiters form beyond the ice line and migrate inwards via high eccentricity migration (HEM), a process which would destabilize the orbits of short-period companions \citep{HEM}. Studies of the Rossiter-Mc\textcolor{black}{L}aughlin effect have also found the fingerprints of high eccentricity migration \citep{albrecht}. However, statistical work has shown that not all hot Jupiters can form in this way \citep{dawson}, so some hot Jupiters may have close-in planets. Additionally, HEM may not exclude nearby, small planets \citep{fogg}.  

In this paper, we present an analysis of the \thisstar\ system \citep[originally announced by][]{hellier} that was recently observed by the \Kepler\ Space Telescope in its new K2 operating mode \citep{howell}. In addition to the previously known hot Jupiter in a 4.16-day orbit, the K2 data reveal two more transiting planets: a super-Earth in a 19-hour orbit, and a Neptune-sized planet in a 9-day orbit. We process the K2 data, determine the planetary properties, and measure the transit times of the three planets.  We find that the measured TTVs are consistent with the theoretical TTVs expected from this system and measure or place limits on the planets' masses. Finally, we perform many dynamical simulations of the \thisstar\ system to assess its stability. 

\section{K2 Data}\label{k2data}

\begin{figure*}[t!]
\epsscale{1}
  \begin{center}
      \leavevmode
\plotone{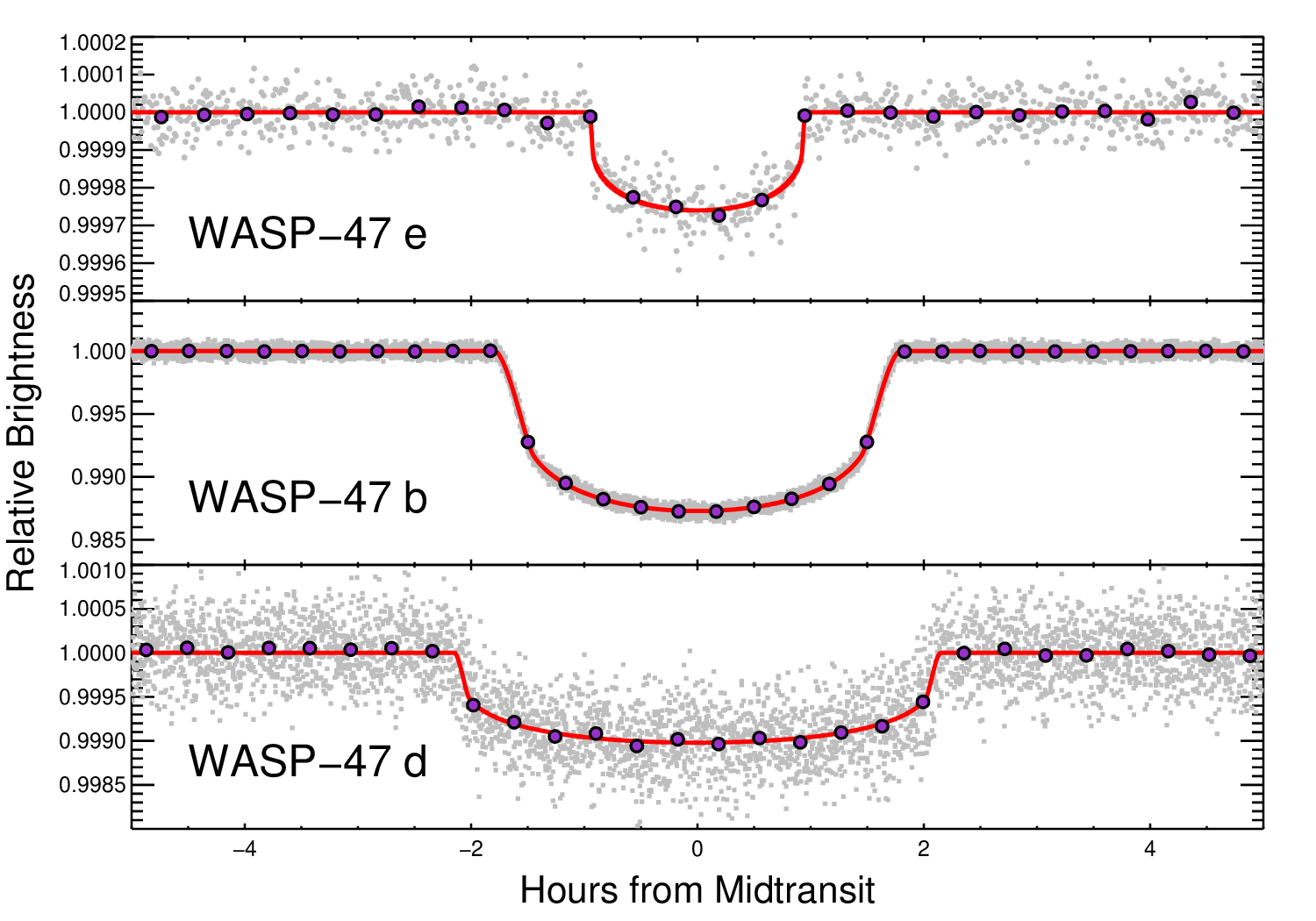}
\caption{Phase-folded short cadence K2 light curve overlaid with our best-fit transit model (red curves), and binned points (purple circles). In the top panel (\thissecondplanet), the grey circles are bins of roughly 30 seconds. In the middle and bottom panels (\thisplanet\ and \thisthirdplanet), the grey squares are the individual K2 short cadence datapoints.} \label{transitfits}
\end{center}
\end{figure*}

\Kepler\ observed K2 Field 3 for 69 days between 14 November 2014 and 23 January 2015. After the data were publicly released, one of us (HMS) identified additional transits by visual inspection of the Pre-search Data Conditioned (PDC) light curve of \thisstar\ (designated EPIC 206103150) produced by the \Kepler/K2 pipeline. We confirmed the additional transits by analyzing the K2 pixel level data following \citet{vj14}. A Box-Least-Squares \citep[BLS; ][]{kovacs} periodogram search \textcolor{black}{\citep[as implemented in][]{philm}} of the processed long cadence light curve identified the 4.16-day period hot Jupiter (\thisplanet), a Neptune sized planet in a 9.03-day period (\thisthirdplanet), and a super-Earth in a 0.79-day period (\thissecondplanet). 

Because of the previously known hot Jupiter, \thisstar\ was observed in K2's ``short cadence'' mode, which consists of 58.3 second integrations in addition to the standard 29.41 minute ``long cadence'' integrations. K2 data are dominated by systematic effects caused by the spacecraft's unstable pointing which must be removed in order to produce high quality photometry. We began processing the short cadence data following \cite{vj14} to estimate the correlation between K2's pointing and the measured flux (which we refer to as the K2 flat field). We used the resulting light curve and measured flat field as starting points in a simultaneous fit of the three transit signals, the flat field, and long term photometric variations \citep[following ][]{hip116454}. The three planetary transits were fit with \cite{mandelagol} transit models, the flat field was modeled with a spline in \Kepler's pointing position with knots placed roughly every 0.25 arcseconds, and the long term variations were modeled with a spline in time with knots placed roughly every 0.75 days. We performed the fit using the Levenberg-Marquardt least squares minimization algorithm \citep{mpfit}. The resulting light curve\footnote{The short cadence light curve is available for download at \url{www.cfa.harvard.edu/~avanderb/wasp47sc.csv}} shows no evidence for K2 pointing systematics, and yielded a photometric precision of 350 parts per million (ppm) per 1 minute exposure. For comparison, during its original mission, \Kepler\ also achieved 350 ppm per 1 minute exposure on the equally bright ($\rm{Kp}=11.7$) KOI 279. 

We measured planetary and orbital properties by fitting the short cadence transit light curves of all three planets with \citet{mandelagol} transit models using Markov Chain Monte Carlo (MCMC) algorithm with affine invariant sampling \citep{goodmanweare}. \textcolor{black}{We used 50 walkers and 9000 links, and confirmed convergence with the test of \citet{geweke2} and a comparison of the Gelman-Rubin statistics for each parameter.  We fit for the $q_1$ and $q_2$ limb darkening parameters from \citet{kipping},} and for each planet, we fit for the orbital period, time of transit, orbital inclination, scaled semimajor axis $a/R_\star$, and $R_{p}/R_{\star}$. Our best-fit model is shown in Figure \ref{transitfits} and our best-fit parameters are given in Table \ref{bigtable}. Our measured planetary parameters for \thisplanet\ are consistent with those reported in \citet{hellier}.

\begin{figure}[t!]
\epsscale{1}
  \begin{center}
      \leavevmode
   \includegraphics[width=3.4in]{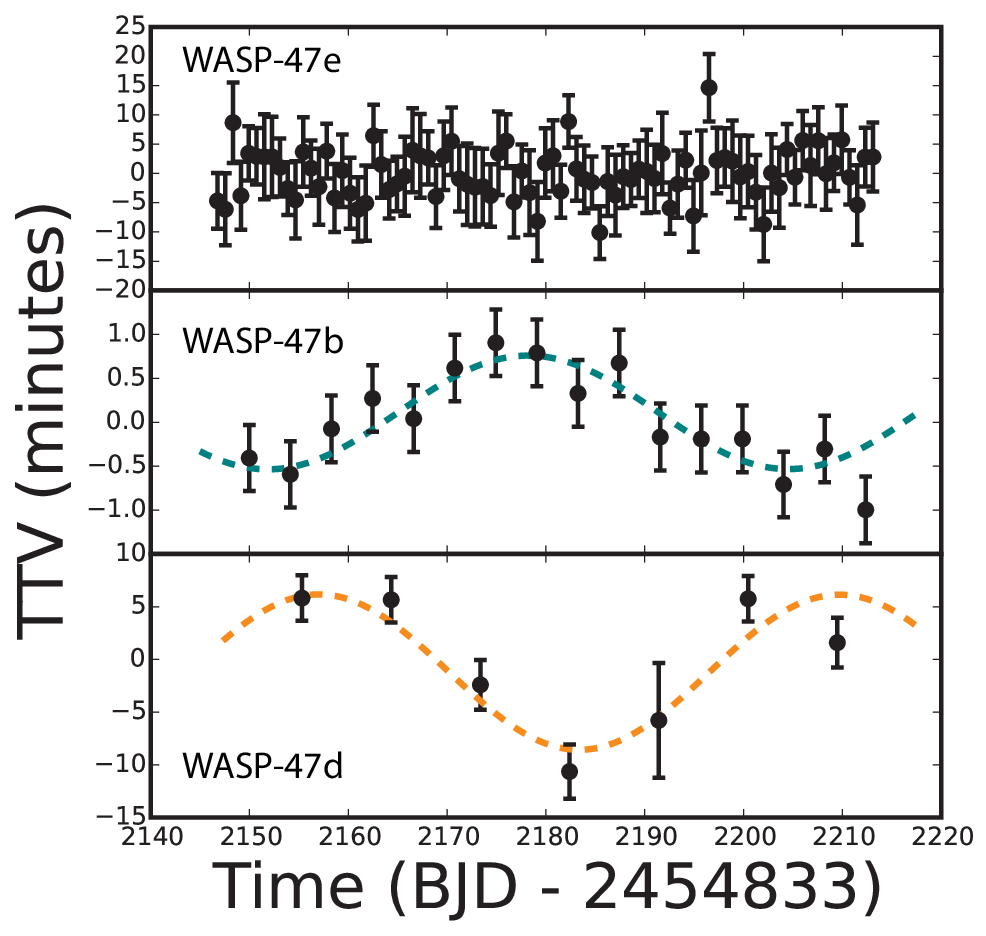}
   \caption{Top: observed TTVs for \thissecondplanet. Middle: observed TTVs for \thisplanet, overlaid (for visual clarity) with a teal sine curve with period equal to the expected 52.67-day super-period. Bottom: Observed TTVs for \thisthirdplanet, overlaid (for visual clarity) with an orange sine curve with the expected super-period. When analyzing the transit times, we did not use the sine fits, they are simply to guide the eye. } \label{ttvs}
\end{center}
\end{figure}

We also fitted for the transit times and transit shapes of each transit event in the short cadence light curve simultaneously (due to the relatively short orbital periods sometimes causing two transits to overlap) using MCMC. 
Our measured transit times\footnote{\textcolor{black}{Tables available at \url{https://github.com/jxcbecker/Data/tree/main/WASP-47}}} are shown in Figure \ref{ttvs}. We find that the TTVs of \thisplanet\ and \thisthirdplanet\ are detected at high significance.  The two TTV curves are anti-correlated and show variations on a timescale of roughly 50 days. This is consistent with the TTV super-period we expect for planets orbiting in this configuration, which we calculate to be $P_{TTV} = 52.67$ days using Equation 7 of \citet{resttvs}.

\section{Validation of \thissecondplanet\ and \thisthirdplanet}

Transiting planet signals like those we find for \thissecondplanet\ and \thisthirdplanet\ can be mimicked by a variety of astrophysical false positive scenarios. In this section, we argue that this is unlikely in the case of the \thisstar\ system.  The hot Jupiter, \thisplanet, was discovered by \citet{hellier} and confirmed with radial velocity (RV) follow-up, which showed no evidence for stellar mass companions or spectral line shape variations, and detected the spectroscopic orbit of the planet. In the K2 data, we detect transit timing variations of \thisplanet\, which are anti-correlated with the transit timing variations of \thisthirdplanet, and which have a super-period consistent with what we expect if both of these objects are planets. The TTVs therefore confirm that \thisthirdplanet\ is a planet in the same system as \thisplanet. \textcolor{black}{We also fitted the transit durations with a power law and found that they followed the expected $P^{1/3}$ relation (when $P$ is orbital period) for planets orbiting a single star.}

The light curve is not of sufficiently high quality to detect TTVs for the smaller \thissecondplanet, so we validate its planetary status statistically. We do this using \vespa\ \citep{morton2015}, an implementation of the procedure described in \citet{morton2012}. Given constraints on background sources which could be the source of the transits, a constraint on the depth of any secondary eclipse, the host star's parameters and location in the sky, and the shape of the transit light curve, \vespa\ calculates the probability of a given transit signal being an astrophysical false positive. Both visual inspection of archival imaging and a lucky imaging search \citep{lucky} found no close companions near \thisstar, but the lucky imaging is not deep enough to rule out background objects that could cause the shallow transits of \thissecondplanet. Following \citet{montet}, we define a conservative radius inside of which background sources could cause the transits. We adopt a radius of 12 arcseconds; we detect the transits with photometric apertures as small as 6 arcseconds in radius and allow for the possibility that stars outside of the aperture could contribute flux due to \Kepler's 6 arcsecond point spread function. We find that \thissecondplanet\ has a false positive probability (FPP) of roughly $5\times10^{-4}$. \textcolor{black}{We find using the expressions from \cite{lissauer} that since this is a three-planet system, its FPP decreases to less than $10^{-5}$. As such,} we consider \thissecondplanet\ to be validated as a \emph{bona fide} planet. 

\section{Dynamical Simulations}\label{stability}

\subsection{Stability Analysis}

We test the dynamical stability of the WASP-47 planetary system with a large ensemble of numerical simulations. The K2 data determine the orbital periods of the three bodies to high precision and place constraints on the other orbital elements. We sample the allowed ranges of the orbital elements for all three planets, randomizing the orbital phases of the three bodies. We assigned masses by sampling the distribution of \citet{wolfgang} for the measured planet radii. We chose eccentricities from a uniform distribution that extends up to $e$=0.3. We discard systems that do not satisfy the stability criteria enumerated in \citet{fabstab}.

Given a set of 1000 such initial conditions, we numerically integrate the systems using the \texttt{Mercury6} integration package \citep{m6}. We use a Bulirsch-Stoer (B-S) integrator, requiring that system energy be conserved to 1 part in $10^9$. We integrate the system for a total simulation length of 10 Myr, unless the system goes unstable on a shorter time scale due to ejection of a planet, planetary collisions, or accretion of a planet by the central star. To perform these computationally intensive simulations, we use the Open Science Grid \citep[OSG;][]{osg1} accessed through XSEDE \citep{xsede1}.

\begin{figure*}[t!]
\epsscale{1}
  \begin{center}
      \leavevmode
   \includegraphics[width=7.5in]{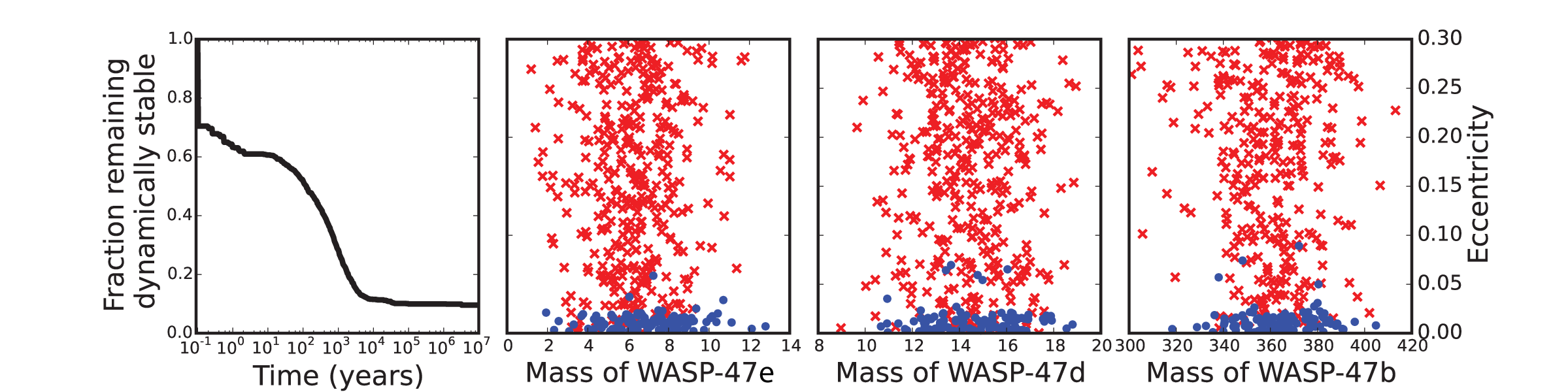} 
\caption{Results from an ensemble of 1000 numerical integrations testing the stability of the system. The left panel shows the fraction of the systems that survive as a function of time. The other panels show the starting mass and eccentricity of the three planets sampled over 1000 trials; the blue points represent systems that are stable, whereas the red crosses depict systems that become dynamically unstable. } 
\label{fig:stability}
\end{center}
\end{figure*}

The results from this numerical survey are shown in Figure \ref{fig:stability}. The left panel shows the fraction of systems remaining stable as a function of time. About 30\% of the systems are unstable over short time scales, and almost 90\% of the systems are unstable over long time scales. Once the systems reach ages of $\sim10^4$ yr, they tend to survive over the next three orders of magnitude in integration time. The remaining three panels show the mass and initial eccentricity of the three planets, sampled from the distributions specified above. One clear trend is that low eccentricity systems tend to survive, whereas systems with $e_{p}>0.05$ are generally unstable. A second trend that emerges from this study is that stability does not depend sensitively on the planet masses (provided that the orbits are nearly circular). Stable systems arise over a wide range of planet masses, essentially the entire range of masses allowed given the measured planetary radii. 

\thisplanet\ and \thisthirdplanet\ orbit within about 20\% of the 2:1 mean motion resonance (MMR). For completeness, we carried out a series of numerical integrations where the system parameters varied over the allowed, stable range described above. In all trials considered, the resonance angles were found to be circulating rather than librating, so there is no indication that the system resides in MMR. 

Each of the numerical integrations considered here spans 10 Myr, which corresponds to nearly one billion orbits of the inner planet. Tidal interactions occur on longer timescales than this and should be considered in future work. In particular, the survival of the inner super-Earth planet over the estimated lifetime of the \thisstar\ system could place limits on the values of the tidal quality parameters $Q$ for the bodies in the system.  

\subsection{Theoretical TTVs}
We performed a second ensemble of numerical simulations to estimate the magnitude of transit timing variations in the WASP-47 system. We used initial conditions similar to those adopted in the previous section, but with starting eccentricities $e<0.1$. 

\begin{figure}[htbp] 
   \centering
   \includegraphics[width=3.4in]{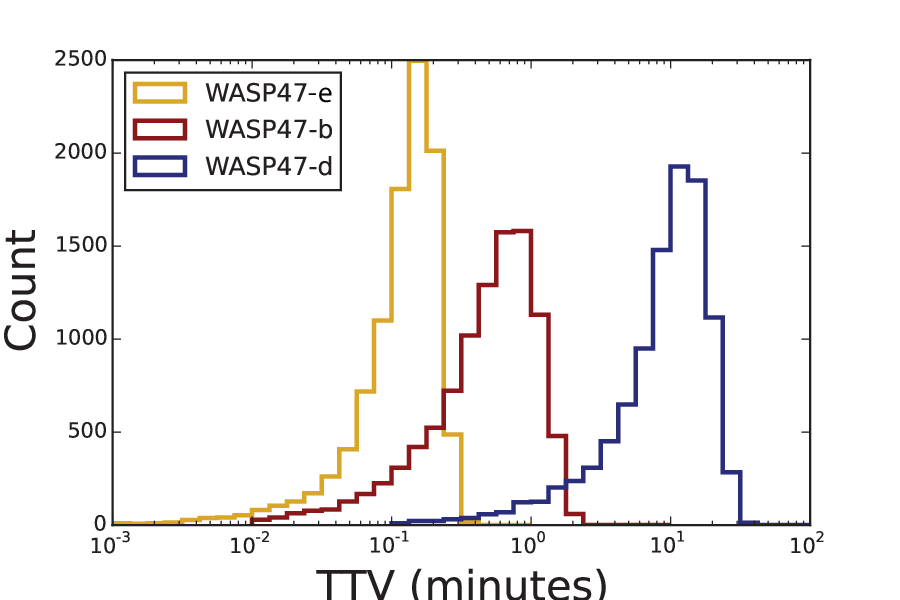} 
   \caption{Histograms of potential transit timing variations for each planet. For a large selection of (likely) dynamically stable initial conditions, we integrated the system forward over a ten-year timescale and extracted the expected TTV amplitude.}
   \label{fig:ttvhist}
\end{figure}

We integrated each realization of the planetary system for 10 years using the \texttt{Mercury6} B-S integrator with time-steps $<0.5$ seconds. We extracted transit times from each integration for each planet, resulting in theoretical TTV curves. The resulting distributions of predicted TTV amplitudes are shown in Figure \ref{fig:ttvhist}. The three distributions have approximately the same shape and exhibit well defined peaks. The TTV amplitudes we measured in Section \ref{k2data} are consistent with the distributions we produced theoretically. 

\subsection{Mass Measurements from the Transit Timing Variations}

\begin{figure}[htbp] 
   \centering
   \includegraphics[width=3.4in]{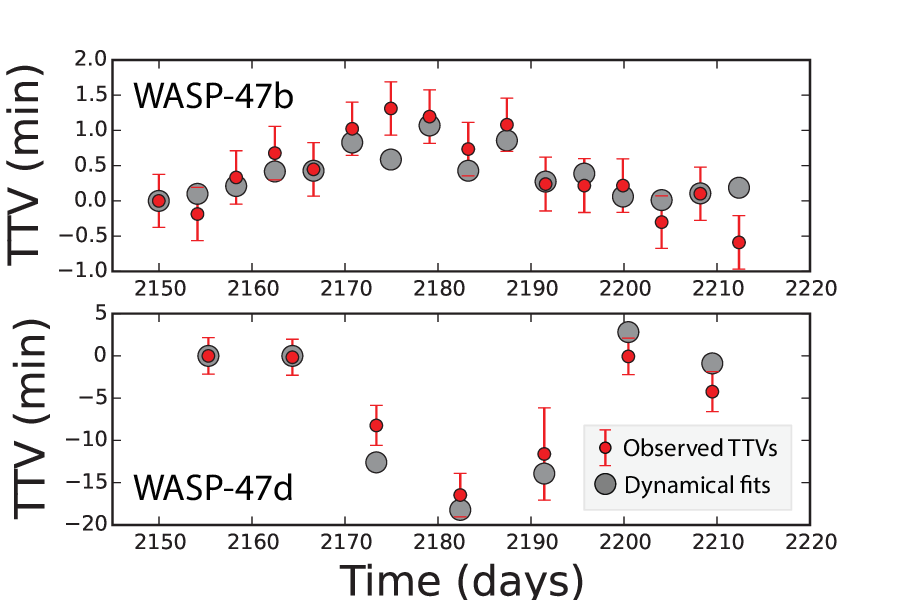} 
   \caption{\textcolor{black}{The best-fit theoretical points (red, with error bars) are overlaid with the observed TTVs (grey circles) for the best fit system parameters given in Table \ref{bigtable}.}}
   \label{ttvplot}
\end{figure}

We measure the TTVs with high enough precision that dynamical fits can give estimates of the planetary masses. We use \texttt{TTVFAST} \citep{ttvfast} to generate model transit times for each observed epoch for each planet, and use \emcee\ \citep{emcee}, an MCMC algorithm with affine invariant sampling, to minimize residuals between the observed TTVs and these model TTVs. In these fits, we allow each planet's mass, eccentricity, argument of pericenter, and \textcolor{black}{time of first transit} to float. We imposed a uniform prior on eccentricity between 0 and \textcolor{black}{0.06} (as required for long-term stability). We initialized the chains with random arguments of pericenter and masses drawn from the \citet{hellier} mass posterior for \thisplanet\, and the distribution of \citet{wolfgang} for \thissecondplanet\ and \thisthirdplanet. We used \textcolor{black}{64} walkers and \textcolor{black}{20000} iterations to explore the parameter space, and discarded the first 2500 iterations as `burn-in'. We confirmed that the MCMC chains had converged using the test of \citet{geweke2} \textcolor{black}{and the Gelman-Rubin statistics (which were below 1.05 for every parameter)}. \textcolor{black}{The best-fit model points are overlaid with the observed TTVs for the outer two planets in Figure \ref{ttvplot}.}

We find that we are able to measure the masses of \thisplanet\ and \thisthirdplanet\, and place an upper limit on the mass of \thissecondplanet. We additionally provide limits on the quantities $e_{c}\cos{\omega_{c}}-e_{b}\cos{\omega_{b}}$. These masses and limits are summarized in Table \ref{bigtable}. We measure a mass of \mplb\umplb\ \mearth\ for \thisplanet, which is consistent with the mass measured by \citet{hellier} of $362\pm16$ \mearth\ at the 1--$\sigma$ level. The mass of \thisthirdplanet\ is \mpld \umpld $\ M_\oplus$.  Only an upper limit can be placed on \thissecondplanet\ of \mplc $M_\oplus$.

\section{Discussion}

\thisstar\ is unusual: it is the first hot Jupiter discovered to have additional, close-in companion planets. Using the Exoplanet Orbit Database \citep{han}, we found that among the 224 systems containing a planet with mass greater than 80 \mearth\ and orbital period less than 10 days, only six contain additional planets, and none of them have additional planets in orbital periods shorter than 100 days. That the additional planets in the \thisstar\ system are coplanar with the hot Jupiter and that the planets are unstable with $e\gtrsim0.05$ implies that the \thisstar\ planets either migrated in a disk or some damping near the end of migration took place to bring them into their present compact architecture. 

\textcolor{black}{The continued existence of the companions in this system indicates that HEM cannot serve as the sole formation mechanism for hot Jupiters. HEM would likely have disrupted the orbits of the smaller planets. It is quite possible that there is more than one potential formation mechanism for hot Jupiters. Additionally, recent observations have identified an additional Jupiter-mass planet in a 571-day orbit (called WASP-47c; Neveu-VanMalle et al. submitted) in this system, making this the first hot Jupiter with both close-in companions and an external perturber. Future dynamical work will place limits on the architecture of this system.}

\thisstar\ is a rare system for which planet masses can be determined using TTVs measured from the K2 data set. This is because (a) the planets are far enough away from resonance that the super-period (52.7 days) is shorter than the K2 observing baseline (69 days), and (b) the planets are massive enough that the TTVs are large enough to be detectable. The detection of TTVs was also aided by the fact that \thisstar\ was observed in short cadence mode, which is unusual for K2. 

Finally, \thisstar\ is a favorable target for future follow-up observations. The V-band magnitude is 11.9, bright enough for precision RV follow-up studies. The K2 light curve shows no evidence for rotational modulation, indicating that \thisstar\ is photometrically quiet and should have little RV jitter. Measuring the mass of the two planetary companions with RVs could both improve the precision of the inferred masses and test the consistency of TTV and RV masses, between which there is some tension \citep[e.g. KOI 94:][]{koi94weiss, koi94masuda}. The 1.3\% depth of the transits of \thisplanet\ makes it easily detectable from the ground. Previous ground based searches for transit timing variations of hot Jupiters have attained timing uncertainties of $\sim20$ seconds, lower than the measured TTV amplitude for \thisplanet\ \citep{gibson}. Follow-up transit observations could place additional constraints on the masses of the \thisstar\ planets. 

\section{Conclusions}

In this work we have studied the \thisstar\ planetary system by using data from the \Kepler/K2 mission along with supporting theoretical calculations. Our main results can be summarized as follows:
\begin{enumerate}
\item In addition to the previously known hot Jupiter companion \thisplanet, the system contains two additional planets that are observed in transit. The inner planet has a ultra-short period of only \perplc\ days, and radius of \rplc\ \rearth. The outer planet has a period of \perpld\ days and a radius of \rpld\ \rearth, comparable to Neptune.
\item The system is dynamically stable. We have run 1000 10 Myr numerical integrations of the system. The planetary system remains stable for the 10 percent of the simulations that start with the lowest orbital eccentricities.
\item The particular planetary system architecture of WASP-47 results in measurable TTVs, which are in good agreement with the TTVs we find from numerical integrations of the system. We use the TTVs to measure the masses of \thisplanet\ (consistent with RV measurements) and \thisthirdplanet.
\item This compact set of planets in nearly circular, coplanar orbits demonstrates that at least a subset of Jupiter-size planets can migrate in close to their host star in a dynamically quiet manner, \textcolor{black}{suggesting that there may be more than one migration mechanism for hot Jupiters}. 

\end{enumerate}

The \thisstar\ planetary system provides a rare opportunity where planets can be both inferred from TTVs and seen in transit. Future observations comparing the system parameters inferred from TTVs with those inferred from RVs will qualitatively test TTVs as a general technique. 
 
\acknowledgements
\textcolor{black}{We are grateful to Kat Deck and Tim Morton for significant assistance. We thank Marion Neveu-VanMalle, Eric Bell, Moiya McTier, Mark Omohundro, and Alexander Venner for useful conversations. We thank John Johnson for his guidance and the anonymous referee for their very useful comments. 
We acknowledge the Planet Hunters team for its community. This work used the Extreme Science and Engineering Discovery Environment (NSF ACI-1053575) and the OSG (NSF, DoE). J.B. and A.V are supported by the NSF GRFP (DGE 1256260, DGE 1144152).
This research has used the Exoplanet Data Explorer at \url{http://www.exoplanets.org}. The data in this paper were obtained from the Mikulski Archive for Space Telescopes. This paper includes data collected by the \Kepler/K2 mission (funding provided by the NASA Science Mission directorate), and we gratefully acknowledge the efforts of the entire \Kepler/K2 team. 
Facilities: \facility{Kepler/K2}}

\begin{deluxetable*}{lcccc}
\tablecaption{System Parameters for \thisstar \label{bigtable}}
\tablewidth{0pt}
\tablehead{
  \colhead{Parameter} & 
  \colhead{Value}     &
  \colhead{} &
  \colhead{68.3\% Confidence}     &
  \colhead{Comment}   \\
  \colhead{} & 
  \colhead{}     &
  \colhead{} &
  \colhead{Interval Width}     &
  \colhead{}  
}
\startdata
\emph{Stellar Parameters} & & & \\
Right Ascension & 22:04:48.7 & & &  \\
Declination & -12:01:08 & & &  \\

$M_\star$~[$M_\odot$] & \mst & $\pm$&$ \umst$ & A \\
$R_\star$~[$R_\odot$] & \rst & $\pm$&$ \urst$ & A \\
\textcolor{black}{Limb darkening $q_1$~} & \textcolor{black}{\ldone}  & \textcolor{black}{$\pm$}&\textcolor{black}{$ \uldone$} & B,D \\
\textcolor{black}{Limb darkening $q_2$~} & \textcolor{black}{\ldtwo}  & \textcolor{black}{$\pm$}&\textcolor{black}{$ \uldtwo$} & B,D \\

$\log g_\star$~[cgs] & \loggst & $\pm$&$ \uloggst$ & A \\
\meh & $\fe$ & $\pm$&$ \ufe$ & A \\
$T_{\rm eff}$ [K] & \teff & $\pm$&$ \uteff$ & A\\
\textcolor{black}{$e_{d}\cos{\omega_{d}}-e_{b}\cos{\omega_{b}}$} &  \textcolor{black}{-0.0002 }&$\pm$  & \textcolor{black}{0.0195}  & A,B,C\\
\textcolor{black}{$e_{d}\sin{\omega_{d}}-e_{b}\sin{\omega_{b}}$} &  \textcolor{black}{0.0039} & $\pm$  & \textcolor{black}{0.0179}   & A,B,C\\
 & & \\
 
\emph{\thisplanet} & & & \\
Orbital Period, $P$~[days] & \perplb & $\pm$&$ \uperplb $ & B \\
Radius Ratio, $(R_P/R_\star)$ & \rprstb & $\pm$&$ \urprstb$ & B \\
Scaled semimajor axis, $a/R_\star$  & \arstb & $\pm$&$ \uarstb$ & B \\
Orbital inclination, $i$~[deg] & \inclb & $\pm$&$ \uinclb$ & B \\
Transit impact parameter, $b$ & \impb & $\pm$&$ \uimpb$ & B \\
Time of Transit $t_{t}$~[BJD] & \ttransitb & $\pm$& \uttransitb & B\\ 
\textcolor{black}{TTV amplitude} \,[min] & 0.63 & $\pm$ & 0.10 & B \\
$M_P$~[\mearth] & \mplb  &   & \umplb & A,B,C \\
$R_P$~[\rearth] & \rplb &   $\pm$&$ \urplb$  & A,B \\
 & & \\
 
\emph{\thissecondplanet} & & & \\
Orbital Period, $P$~[days] & \perplc & $\pm$&$ \uperplc $ & B \\
Radius Ratio, $(R_P/R_\star)$ & \rprstc & $\pm$&$ \urprstc$ & B \\
Scaled semimajor axis, $a/R_\star$  & \arstc & $\pm$&$ \uarstc$ & B \\
Orbital inclination, $i$~[deg] & \inclc & $\pm$&$ \uinclc$ & B \\
Transit impact parameter, $b$ & \impc & $\pm$&$ \uimpc$ & B \\
Time of Transit $t_{t}$~[BJD] & \ttransitc & $\pm$& \uttransitc & B\\ 
\textcolor{black}{TTV amplitude} \,[min] & \multicolumn{3}{c}{\textcolor{black}{$<$1.2 min for any TTV period $<$ 80 days}} & B\\
$M_P$~[\mearth] & \mplc  &   & \umplc  & C \\
$R_P$~[\rearth] & \rplc &   $\pm$&$ \urplc$  & A,B \\
 & & \\

\emph{\thisthirdplanet} & & & \\
Orbital Period, $P$~[days] & \perpld & $\pm$&$ \uperpld $ & B \\
Radius Ratio, $(R_P/R_\star)$ & \rprstd & $\pm$&$ \urprstd$ & B \\
Scaled semimajor axis, $a/R_\star$  & \arstd & $\pm$&$ \uarstd$ & B \\
Orbital inclination, $i$~[deg] & \incld & $\pm$&$ \uincld$ & B \\
Transit impact parameter, $b$ & \impd & $\pm$&$ \uimpd$ & B \\
Time of Transit $t_{t}$~[BJD] & \ttransitd & $\pm$& \uttransitd & B\\ 
\textcolor{black}{TTV amplitude} \,[min] & 7.3 & $\pm$ & 1.9 & B\\
$M_P$~[\mearth] & \mpld &   & \umpld & C \\
$R_P$~[\rearth] & \rpld &   $\pm$&$ \urpld$  & A,B \\
 & & \\

\enddata

\tablecomments{A: Parameters come from \citet{mortier}. B: Parameters come from analysis of the K2 light curve. C. Parameters come from dynamical fits to the observed transit timing variations. \textcolor{black}{D: We report the magnitude of the impact parameter, whereas the true value could be positive or negative.}}

\end{deluxetable*}

\end{document}